\documentstyle[11pt,bezier,epsf,cite]{article}
\author{
   Yu.Holovatch, T. Yavors'kii}
\author{
   {\sc Yu.~Holovatch$^{(1)}$,
   T.~Yavors'kii$^{(2)}$},
   \\[1.5ex]
  \it $^{(1)}$Institute for Condensed Matter Physics
   \\
   \it of the Ukrainian Academy of Sciences
   \\
   \it 1~Svientsitskii St., UA--290011 Lviv, Ukraine
   \\
\it $^{(2)}$Chair for Theoretical Physics,
\\
\it Ivan Franko Lviv State University
\\
\it 12~Drahomanov str., UA--290005 Lviv, Ukraine}

\title{\Large\bf
CRITICAL EXPONENTS OF THE DILUTED ISING MODEL BETWEEN DIMENSIONS 2 AND
   4.  }
\begin{document}
\maketitle
\begin{abstract}
{\small
Within the massive field theoretical renormalization group
ap\-pro\-ach the ex\-pres\-si\-ons for the $\beta$- and $\gamma$-
functions of the an\-iso\-tro\-pic $mn$-vector model are obtained
for general space dimension $d$
in three-loop approximation.
Resumming corresponding asymptotic series,
cri\-ti\-cal exponents for the case of the weakly diluted quenched Ising
model ($m=1$, $n=0$), as well as estimates for the marginal order parameter
component number $m_c$ of the weakly diluted quenched $m$-vector model
are calculated as func\-ti\-ons of $d$ in the region
$2 \leq d <4$.  Conclusions concerning the effectiveness of different
resummation techniques are drawn. }
\end{abstract}



{\it To appear in J.Stat.Phys. vol. 92, Nos 5/6}

\newpage

\section*{Introduction}
Study of the critical behaviour of the Ising model has several
attractions. On the one hand, the Ising-like models are simple enough, which
is of a special advantage in the statistical physics. On the other hand, in
spite of their simplicity, such models show rich and interesting behaviour at
the critical point. Also, the existence of the exact solution for the
two-dimensional Ising model often makes it an object for verifying
different approximation schemes. All the
stated above yielded the high interest devoted to the problem. In particular
a great deal of generalization of the model appeared.
Among different ways of generalization, much attention has been
devoted to the affect of the impurities on the critical behaviour of the
Ising-like models as well as to the investigation of critical regimes of the
models on the lattices of a non-integer dimension ($d$).
There have been devised different realizations of the last stated
generalization.  For example, one can approach the concept of non-integer
dimensionality either by explicit construction of the non-integer dimensional
object, which leads to the concept of a fractal \cite{Mandelbrot'77}, or by
formal carrying out an analytic continuation of the function, which by
definition depends on a natural value of dimension.

Within the theory of critical phenomena the latter ambiguity was reflected in
examining the critical behaviour of the many-particle systems on fractal
\cite{GefenMandelbrot'80,GefenAharony'81} or on abstract hypercubic
lattices of the non-integer $d$. There arosed a question whether a model on
a fractal lattice (being scale invariant) possesses universality as well as a
system on a hypercubic lattice (having translation invariance). The problem
has been widely studied but still remains open
\cite{BonnierLeroyerMeyers'88,JezewskiTomczak'91a,JezewskiTomczak'91b,Jezewski'94}.
Today's point of view states that the usual demand for ''strong
universality'' (in sense of critical properties depending only on symmetry
of the order parameter, interaction range and space dimension) seems not to
be obeyed by fractal lattice systems, and for them the concept of
universality itself should be revised
\cite{Hu'86,WuHu'87}.

Speaking about the study of Ising-like models on analytically continued
hypercubic lattices of non-integer $d$, one should note a great variety of
theoretical approaches devised for these problems. These include:
the
Wilson-Fisher $\epsilon$-expansion \cite{Wilson'72} improved by the summation
method \cite{GuillouJustin'87};
Kadanoff lower-bound renormalization applied
to some special non-integer dimensions \cite{KatzDrotz'77};
high-temperature
expansion improved by a variation technique \cite{BonnierHontebeurie'91};
finite-size scaling method applied to numerical transfer-matrices data
\cite{Novotny'92a,Novotny'92b};
new perturbation theory based on the physical branch of the solution of the
renormalization group equation
\cite{FilippovRadievskii'92,Filippov'92a,Filippov'92b,BreusFilippov'93};
fixed dimension renormalization group technique
\cite{Parisi'73,Parisi'80}
applied directly to arbitrary non-integer $d$
\cite{Holovatch'93,HolovatchKrokhmal's'kii'94}.
Perhaps the first paper devoted to the study of the Ising model in different,
however not non-integer dimension, was
\cite{FisherGaunt'64}
where non-universal properties of the model were discussed.
All these approaches, as well as the computer simulations, confirm the
correctness of the universality hypothesis also for non-integer $d$
hypercubic lattices and allow to obtain the critical exponents as
functions of $d$ with high accuracy.

In spite of the variety of approaches to treat the general non-integer
$d$ case
\cite{
GuillouJustin'87,KatzDrotz'77,BonnierHontebeurie'91,Novotny'92a,
Novotny'92b,FilippovRadievskii'92,Filippov'92a,Filippov'92b,
BreusFilippov'93,Holovatch'93,HolovatchKrokhmal's'kii'94}
the results for the Ising model critical exponents obtained on their basis
lay close to each other; the above mentioned analytic continuations
may appear actually equivalent.
Note, however, that the analytic continuation in $\varepsilon=4-d$ at general
non-integer dimension leads to the fail of the Yang-Lee theorem
\cite{BakerBenofy'82}.
On the other hand, the study of Ising-like spin systems on non-integer dimension
hypercubic lattices cannot be reduced to the fractal lattice systems
\cite{BakerBenofy'82}
except for the case of vanishing lacunarity limit
\cite{BonnierLeroyerMeyers'88}
and thus is the task of individual interest.

Returning to the study of the critical behaviour at integer $d$, one should
note that the problem becomes more complicated when studying
spin systems with a structural disorder. Whereas the case of the annealed
disorder is of less interest from the point of view of determining
asymptotical values of critical exponents \cite{footnote1}, the weak {\bf
quenched} disorder has been a subject of intensive study. Here the Harris
criterion \cite{Harris'74} has been devised.  It states that if the heat
capacity exponent $\alpha_{pure}$ of a pure model is negative, that is the
heat capacity has no divergence at the critical point, impurities do not
affect the critical behaviour of the model in the sense that critical
exponents remain unchanged under dilution.  Only in the case $\alpha_{pure} > 0$,
the critical behaviour of the disordered model is governed by a
{\em new set of critical exponents}. As far as for a $3d$ $m$-vector spin
model only the $3d$ Ising model ($m=1$) is characterized by
$\alpha_{pure}>0$, it is the Ising model which is of special interest. And
because of the triviality of the annealed disorder in the sense mentioned
above, the most interesting object for study is just the quenched Ising
model.  The appearance of a set of new critical exponents for that model at
$d=3$ is confirmed by the experiments
\cite{BirgeneauCowleyShirane'83,MitchellCowleyYoshizawa'86,ThurstonPetersBirgeneau'88},
renormalization group $(RG)$
calculations
\cite{Lubensky'75,Khmelnitskii'75,GrinsteinLuther'76,SokolovShalaev'81,NewmanRiedel'82,Jug'83,MayerSokolov'84,MayerSokolovShalaev'89,Mayer'89,JanssenOerdingSengespeick'95},
Monte-Carlo $(MC)$
\cite{ChowdhuryStaufer'86,MarroLabartaTejada'86,WangWohlert'90,Heuer'90,Heuer'93}
and $MCRG$
\cite{HoleyFahnle'90} simulations.

The situation is not so simple for the $2d$
Ising model. Onsager exact solution of the pure model proves the
logarithmic divergence of heat capacity, which yields $\alpha_{pure}=0$, and
allows one, in accordance with the Harris criterion, to clasify this case as
a marginal one. Most of the theoretical works suggest that the $2d$ Ising
model with a quenched disorder has the same critical behaviour as the $2d$
pure Ising model (except for logarithmic corrections)
\cite{DotsenkoDotsenko'82,Shankar'87,Shalaev'88,Ludwig'88,Shalaev'89,Ludwig'90,MayerSokolovShalaev'89,Mayer'89}
(see also review \cite{Shalaev'94}). This result is corroborated
by $MC$-simulations on two-dimensional lattices
\cite{WangSelkeDotsenko'90,AndreichenkoDotsenkoSelke'90,Heuer'92,TalapovShchur'94,SelkeShchurTalapov}
and experiments
\cite{CowleyBirgeneauShirane'80,HagenCowley'87}.

Deviations from the expected critical exponents, which sometimes are observed
during such computations, are explained by a system being not in the
asymptotic region (see \cite{SelkeShchurTalapov} for recent study).
Nevertheless, some authors assert that for the $2d$ Ising model with
a quenched disorder a new critical behaviour appears
\cite{DotsenkoDotsenko'83,KimPatrascioiu'94}

While the undiluted Ising model at non-integer $d$ was a subject of intensive
study \cite{Wilson'72,GuillouJustin'87,KatzDrotz'77,BonnierHontebeurie'91,Novotny'92a,Novotny'92b,FilippovRadievskii'92,Filippov'92a,Filippov'92b,BreusFilippov'93},
it is not the case for the diluted Ising model. Only the work
\cite{NewmanRiedel'82} can be mentioned here, where the model was studied
within the Golner-Riedel scaling field \cite{GolnerRiedel'75} approach. It is
worthwhile to note that the $\varepsilon$-expansion technique applied to this
model, due to the fact that $RG$-equations appear to be degenerated on the
one loop level, results in $\sqrt{\varepsilon}$-expansion for the critical
exponents \cite{GrinsteinLuther'76}.  The latter is known up to the
three-loop order
\cite{Shalaev'77,JayaprakashKatz'77}.
The equations of the massive field theory at fixed integer $d$
\cite{Parisi'73,Parisi'80}
first applied to the diluted Ising model at $d=2,3$ in
\cite{SokolovShalaev'81,Jug'83} were found to be the most effective method
for investigating this problem.  In order to consider an arbitrary
non-integer $d$ the Parisi approach \cite{Parisi'73,Parisi'80} was
generalized in \cite{HolovatchShpot'92} where critical behaviour of the
model was studied in a two-loop approximation.  The aim of the present work,
based on the massive field theoretical approach, is to make a more
detailed investigation of the critical behaviour of the diluted $O(m)$-vector
model at arbitrary $d$. Though it is the case $m=1$ in which we are
interested most of all, we consider the $RG$-equations for any $m$, which
also allow us to study the crossover in the model at any $d$. We will obtain
the $RG$-equations within the 3-loop approximation and apply to
their analysis different resummation procedures in order to find the most
reliable one.

The set-up of the article is as follows. In the next Section we introduce
the model and the notation. Then we describe the $RG$-procedure adopted
here and give the series for the $RG$-functions of the weakly diluted
quenched $m$-vector model in the three-loop approximation.
Being asymptotic, these series are to be resummed. This is done in Section 2
where different ways of resummation are used.
Section 3 concludes our study giving results for the quantitative
characteristics of the critical behaviour and discussing them.
In the Conclusions we give some general comments to the present work.
In the Appendix we list some lengthy expressions for the coefficients of the
$RG$-functions in the three-loop approximation.

%%%%%%%%%%%%%%%%%%%%%%%%%%%%%%%%%%%%%%%%%%%%%%%%%%%%%%%%%%%%%%%%%%%%%%%%%%%
%                               Section 1.
%                    The Model and the RG - procedure
%%%%%%%%%%%%%%%%%%%%%%%%%%%%%%%%%%%%%%%%%%%%%%%%%%%%%%%%%%%%%%%%%%%%%%%%%%%
\section{The Model and the $RG$ - procedure}

As it is well known, the critical behaviour of the quenched weakly-diluted
$m$~-~vector model is governed by a Lagrangian with two coupling constants
\cite{GrinsteinLuther'76}:
\begin{eqnarray} \label{Lagrangian}
{\cal L}(\phi)&=&\int {\rm d}^dR \Big\{
{1\over 2} \sum_{\alpha=1}^{n} \left[|\nabla \vec{\phi}^\alpha|^2+
m_0^2 |\vec{\phi}^\alpha|^2\right]+ {v_{0}\over 4!}
\left(\sum_{\alpha=1}^{n}|\vec{\phi}^\alpha|^2 \right)^2 +
\nonumber
\\ \label{2.1}
&& {u_{0}\over 4!}
\sum_{\alpha=1}^{n}\left(|\vec{\phi}^\alpha|^2 \right)^2 \Big\},
\end{eqnarray}
in replica limit $n \rightarrow 0$.
Here any $\vec{\phi}^\alpha$ is a $m$-component vector
\\$\vec{\phi}^\alpha=
(\phi^{\alpha,1},\phi^{\alpha,2},\dots,\phi^{\alpha,m})$;
$u_{0} > 0, v_{0} < 0$ are bare coupling constants; $m_0$ is bare mass.

As it was already stated above, we adopt here the massive
field theory renormalization
scheme \cite{Parisi'73,Parisi'80} in order to extract the
critical behaviour governed by (\ref{Lagrangian}). We start from the defined
by (\ref{Lagrangian}) unrenormalized one-particle irreducible vertex
functions
\begin{equation}
\Gamma^{(L,N)}(q_1,..,q_L;p_1,..,p_N;m_0,u_0,v_0;\Lambda_0;d)
\end{equation}
depending on the wave vectors $\{q\}, \{p\}$, bare parameters $m_0, u_0, v_0$
and the ultraviolet momentum cutoff $\Lambda_0$. The vertex functions'
dependence on the space dimension $d$ is explicitly noted here as well. We impose
the renormalization conditions at zero external momenta and non-zero mass
(see \cite{BrezinLeGuillouZinnJustin'76,Amit'84} for instance) at the limit
$\Lambda_0\rightarrow\infty$ for the renormalized functions \cite{footnote2}
$\Gamma^{(0,2)}_R, \Gamma^{(0,4)}_{R,u}, \Gamma^{(0,4)}_{R,v},
\Gamma^{(1,2)}_R$:

\begin{eqnarray}
\Gamma^{(0,2)}_R(p,-p;m,u,v;d)|_{p=0} &=& m^2, \\
\frac{d}{dp^2}\Gamma^{(0,2)}_R(p,-p;m,u,v;d)|_{p=0} &=& 1, \\
\Gamma^{(0,4)}_{R,u}(\{p_i\};m,u,v;d)|_{\{p_i\}=0} &=& m^{4-d}u, \\
\Gamma^{(0,4)}_{R,v}(\{p_i\};m,u,v;d)|_{\{p_i\}=0} &=& m^{4-d}v, \\
\Gamma^{(1,2)}_{R,u}(q;p,-p;m,u,v;d)|_{q=p=0} &=& 1,
\end{eqnarray}
with $m, u, v$ being the re\-nor\-ma\-liz\-ed mass
$m=Z_3m_1=Z_3\Gamma^{(0,2)}(0;m_0,u_0,v_0)$
and couplings $u=m^{d-4}Z^2_3Z^{-1}_{1,u}u_0, v=m^{d-4}Z^2_3Z^{-1}_{1,v}v_0$.
From these conditions there follow expansions for the re\-nor\-ma\-li\-zed
constants for field $(Z_3)$, vertices $u$ $(Z_{1,u})$, $v$ $(Z_{1,v})$ and
$\phi^2$ insertion $(Z_2)$. Subsequently, these define the
coefficients $\beta, \gamma$ entering the corresponding Callan-Symanzik
equation:
\begin{eqnarray}
\beta_u(u,v) &=& \frac{\partial u}{\partial \ln m}|_{u_0, v_0}, \\
\beta_v(u,v) &=& \frac{\partial v}{\partial \ln m}|_{u_0, v_0}, \\
\gamma_{\phi} \equiv \gamma_3 &=& \frac{\partial Z_3}{\partial \ln m}
|_{u_0, v_0}, \\
\bar{\gamma}_{\phi^2} \equiv \gamma_2 &=& -\frac{\partial Z_2}{\partial \ln m}
|_{u_0, v_0}.
\end{eqnarray}

In the stable fixed point $\{u^*, v^*\}$ to be defined by simultaneous
zero of both $\beta$-functions:
\begin{eqnarray} \nonumber
\beta_u(u^*,v^*) &=& 0, \\
\beta_v(u^*,v^*) &=& 0,
\label{fixedpoint}
\end{eqnarray}
the $\gamma_{\phi}$-function gives the critical exponent $\eta$ of
the pair correlation function:
\begin{equation}
\gamma_{\phi}(u^*,v^*) = \eta.
\end{equation}

The correlation length critical exponent $\nu$ is defined in the stable
fixed point by:

\begin{equation}
\bar{\gamma}_{\phi^2}(u^*,v^*) = 2 - \nu^{-1} -
\gamma_{\phi}(u^*,v^*).
\end{equation}

Using familiar scaling relations, one can easily calculate any other
critical exponents on the base of $\eta$ and $\nu$.

Applying the described above procedure,
one obtains in the three-loop approximation \cite{Holovatch'unpubl}
$\beta$- and $\gamma$-functions in the form \cite{footnote3}:
\begin{eqnarray} \label{beta_u}
\beta_u(u,v) & = & -(4-d) u \Big \{ 1 - u - \frac{12}{mn+8} v +
\frac{8}{(m+8)^2} \times
\nonumber\\
&&\Big [ (5m+22)(i_1-\frac{1}{2})+(m+2) i_2 \Big ] u^2 +
\frac{96}{(m+8)(mn+8)} \times
\nonumber\\
&&\Big [ (m+5)(i_1-\frac{1}{2})+ \frac{m+2}{6}
i_2 \Big ] uv + \frac{24}{(mn+8)^2} \times
\nonumber\\
&&\Big [(mn+14)(i_1-\frac{1}{2}) + \frac{mn+2}{3}
i_2 \Big ] v^2 + \beta_u^{(3LA)} + \dots \Big \}
\\ \label{beta_v}
\beta_v(u,v) & = & -(4-d) v \Big \{ 1 - v - \frac{2(m+2)}{m+8} u +
\frac{8}{(mn+8)^2} \times
\\
&&\Big [ (5mn+22)(i_1-\frac{1}{2})+(mn+2) i_2 \Big ] v^2
+ \frac{96(m+2)}{(m+8)(mn+8)} \times
\nonumber\\
&&\Big [ i_1-\frac{1}{2} + \frac{i_2}{6}\Big ] uv +
\frac{24(m+2)}{(m+8)^2} \Big [i_1-\frac{1}{2} +
\frac{i_2}{3} \Big ] u^2 +
\beta_v^{(3LA)} + \dots \Big \}
\nonumber\\
\gamma_{\phi}(u,v) & = & - 2 (4-d) \Big \{
\Big [ \frac{2(m+2)}{(m+8)^{2}} u^{2} + \frac{4(m+2)}{(m+8)(mn+8)} uv +
\nonumber\\
&& \frac{2(mn+2)}{(mn+8)^{2}} v^{2}
\Big ] i_2 +
\gamma_{\phi}^{(3LA)} + \dots \Big \}
\\ \label{gamma}
\bar{\gamma}_{\phi^2}(u,v) & = &  (4-d) \Big \{
\frac{m+2}{m+8} u + \frac{mn+2}{mn+8} v -
12 \Big [ \frac{m+2}{(m+8)^2} u^2 +
\\
&&
\frac{2(m+2)}{(m+8)(mn+8)} uv +
\frac{mn+2}{(mn+8)^2} v^2
\Big ] (i_1-\frac{1}{2}) + \bar{\gamma}_{\phi^2}^{(3LA)} + \dots \Big \}
\nonumber
\end{eqnarray}
Here $d$ is the space dimension, $m$ is the order parameter component number,
$n$ is the replica index, $i_1$ and $i_2$ are dimensionally dependent two-loop
integrals. The corresponding coefficients for three-loop parts are listed in
the Appendix.
The values for the three-loop integrals $i_3 \dots i_8$ which appear in
three-loop coefficients for integer $d=2, 3$ are listed in \cite{NickelMeironBaker'77}.
In particular, substituting loop integrals $i_1, i_2$ as well as
$i_3,\dots,i_8$ in (\ref{beta_u})-(\ref{gamma}) by their values at $d=3$ we get at
$n=0, m=1$ the corresponding functions of the $3d$ weakly diluted Ising
model, which in the 3-loop approximation were obtained in
\cite{SokolovShalaev'81}.  At $d=3$, $m,n$- arbitrary corresponding
expressions coincide with those, obtained for the $3d$ anisotropic
$mn$-vector model in \cite{Shpot'88}. Our idea is to keep the dimensional
dependence of the loop integrals and, being based on their numerical values
for arbitrary $d$ \cite{HolovatchKrokhmal's'kii'94}, to study the
$O(mn)$-model at arbitrary (non-integer) $d$ as well. But for the reason
explained above, the point of main interest here will be the replica limit
$n=0$ of the anisotropic $mn$-vector model, especially the case $m=1$.

Expressions for $\beta$- and $\gamma$-functions will be the starting point
for the qualitative study of the main features of the critical behaviour which
will be done in the next section.

%%%%%%%%%%%%%%%%%%%%%%%%%%%%%%%%%%%%%%%%%%%%%%%%%%%%%%%%%%%%%%%%%%%%%%%%%%%
%                               Section 2.
%                            The Resummation
%%%%%%%%%%%%%%%%%%%%%%%%%%%%%%%%%%%%%%%%%%%%%%%%%%%%%%%%%%%%%%%%%%%%%%%%%%%

\section{The Resummation}

As we have already mentioned, the values of the $\gamma$-functions in a
fixed point $\left(u^{*},v^{*}\right)$ lead to the values of the critical
exponents $\eta$ and $\nu$. However, it is well known now that the
series for $RG$-functions are of
asymptotic nature
\cite{Lipatov'77,LeGuillouBrezinZinnJustin'77,BrezinParisi'78} and
imply the corresponding resummation procedure to extract reliable data on
their basis. Let us note, however, that as to our knowledge the
asymptotic nature of the series for $RG$-functions have been proved
only for the case of the model with one coupling \cite{Eckmann'75}, and
the application of a resummation procedure to the case of several
coupling constants is based rather on general belief than on a proved
fact. One of the resummation procedures, which in different modifications is
most commonly used in the studies of asymptotic series, is known as the
integral Borel transformation \cite{Hardy'48}. However, this technique
implies explicit knowledge of the general term of a series and thus cannot
be applied here, where only truncated sums of the series are known. To get
over this obstacle one represents the so-called Borel-Leroy image of the
initial sum in the form of a rational approximant and in such a way
reconstitutes the general term of the series. The technique which involves a
rational approximation and the Borel transformation together, is known as the
Pad\'e-Borel resummation technique (in the field-theoretical $RG$ content see
\cite{BakerNickelGreen'76,BakerNickelMeiron'78} as an example of its
application).

Note here that the resummation technique, based on the conformal
mapping, which is widely used in the theory of critical phenomena
\cite{LeGuillouZinnJustin'80}, cannot be applied in our case because its
application postulates information on the high order behaviour of the
series for $\beta$- and $\gamma$-functions. The latter is still unknown for
the theory with the Lagrangian (\ref{Lagrangian}).

To summarize up the stated let us write that the Pad\'e-Borel resummation
is performed as follows:
\begin{itemize}
\item
constructing the Borel-Leroy image of the initial sum $S$ of $n$ terms:
\begin{equation} \label{Leroy}
S=\sum_{i=0}^na_ix^i\Rightarrow
\sum_{i=0}^n\frac{a_i(xt)^i}{\Gamma(i+p+1)},
\end{equation}
where $\Gamma(x)$ is the Euler's gamma function and $p$ is an arbitrary
non-negative number. The special cases $p=0$ and $p=1$ correspond to
resumming $\beta$-functions without or with prefactors $u$ and $v$ in
accordance with the structure of the functions (\ref{beta_u})-(\ref{beta_v});
\item
the Borel-Leroy image (\ref{Leroy}) is extrapolated by a rational
ap\-pro\-xi\-mant \linebreak $\left[ M/N \right] (xt)$,
where by $\left[ M/N \right] $ one means the
quotient of two polynomials; $M$ is the order of the numerator and
$N$ is that of the denominator;
\item
the resummed function $S^{res}$ is
obtained in the form:
\begin{equation} \label{p}
S^{res}=\int_0^\infty dt
\exp (-t)t^p\left[ M/N \right] (xt).
\end{equation}
\end{itemize}

In the two variables case only the first step is changed; namely, here we
define the Borel-Leroy image as
\begin{equation}
\sum_{0\leq i+j \leq n}a_{i,j}x^iy^j\Rightarrow
\sum_{0\leq i+j \leq n}\frac{a_{i,j}(xt)^i(yt)^j}{\Gamma(i+j+p+1)}.
\end{equation}
Generalization to the many-variable case is trivial.

Now one can take into account that the second step of the stated scheme can
be done in different ways. One can write down various Pad\'e approximants in
the variable $t$ to obtain within the three-loop approximation the
expressions of the structure $\left[ 2/1 \right]$, $\left[ 1/2 \right]$ and
$\left[ 0/3 \right]$. It is also possible to use Chisholm approximants
\cite{Chisholm'73} in the variables $u$ and $v$, which, generally speaking,
in the same number of loops can be of type $\left[ 3/1 \right]$, $\left[ 2/2
\right]$, $\left[ 1/3 \right]$ and $\left[ 0/4 \right]$, but the
explicit definition of any approximant needs some additional equations now
\cite{Chisholm'73}. The technique, which involves Chisholm approximation
together with the integral Borel transformation is referred to as the
Chisholm-Borel resummation technique. To be consistent, one would have to
apply the all different resummation frameworks in order to obtain reliable
results on their basis and find which of the methods is the most effective.
However, strong restriction on the number of choices can be imposed.

First of all,
an approximant should be chosen in the form reconstituting the
sign-alternating high-order behaviour of the general term of $\beta$- and
$\gamma$-functions, which was confirmed in the particular case $m=1$, $n=2$
and $n=3$ \cite{KleinertThoms'95}.
The approximant generating a sign-alternating series might be chosen in a
form $\left[ M/1 \right]$ with the positive coefficients at the variable $t$
(or $u$ and $v$). Choosing an approximant with a non-linear denominator,
generally speaking, one does not ensure the desired properties.
Direct calculations affirm the argumentation:  $\beta$-functions,
resummed with the Pad\'e-Borel and the Chisholm-Borel methods with
approximants $\left[ M/N \right], N>1$, for $u<0,v>0$ give the roots which
lie far from the expected values which for $d=3$ are known up to the order of
four loops \cite{MayerSokolovShalaev'89} and for general $d$ were calculated
from the two-loop $\beta$-functions \cite{HolovatchShpot'92}.  This is true
for any $p$. The stated results permit us to eliminate from the consideration
approximants with a non-linear denominator.

Note as well that choosing representation of the $RG$-functions (\ref{beta_u})-(\ref{gamma})
in the form of Pad\'e or Chisholm approximant of type $\left[ M/1
\right]$ might result in the appearance of a pole in the expression. Here we
use an analytical continuation of the resulting expressions by evaluating the
principal value of the integral. Treating the task in this way one notes that
the topological structure of the lines of zeros for the resummed by the
Pad\'e-Borel technique $\beta$-functions is very different in the region near
the solution for the mixed fixed point and strongly irregular when passing
through the number of loops. In particular this yields that in the three-loop
approximation there exist two solutions close to the expected value of the
mixed fixed point. To compare, the results obtained within the frames of the
Chisholm-Borel method do not have these faults and are more stable from the
point of view of proceeding in number of loops.

So, the results given below are obtained by the Chisholm-Borel method applied
to the approximant of type $\left[ 3/1 \right]$. In order to determine the
form of this approximant completely one must define two additional
conditions. The approximants are expected to be symmetric in variables $u$
and $v$, otherwise the properties of the symmetry related to these variables
would depend, except for the properties of the Lagrangian, on the method of
calculation.  By the substitution $v=0$ all the
equations which describe the critical behaviour of the diluted model are
converted into appropriate equations of the pure model.  However, if pure
model is solved independently, the resummation technique with the application
of Pad\'e approximant is used.  Thus, Chisholm approximant is to be chosen in
such a way that, by putting any of $u$ or $v$ equal to zero, one obtains
Pad\'e approximant for a one-variable case. This also implies a special
choice of additional conditions.  In the present study amidst all the
possible expressions which satisfy the stated demand we choose Chisholm
approximant $\left[ 3/1 \right]$ by putting coefficients at $u^3$ and $v^3$
to be equal to zero.

%%%%%%%%%%%%%%%%%%%%%%%%%%%%%%%%%%%%%%%%%%%%%%%%%%%%%%%%%%%%%%%%%%%%%%%%%%%
%                               Section 3.
%                                Results
%%%%%%%%%%%%%%%%%%%%%%%%%%%%%%%%%%%%%%%%%%%%%%%%%%%%%%%%%%%%%%%%%%%%%%%%%%%

\section{Results}

Now we are going to apply the mathematical framework which was
discussed in previous sections in order to obtain numerical
characteristics of the critical behaviour of the weakly-diluted Ising model in
general dimensions.
It was noted in the Section 1 that the critical behaviour of the quenched
weakly-diluted Ising model is described by the effective Lagrangian
(\ref{Lagrangian}) in the case $m=1$ and zero replica limit $n=0$. Namely, the
task in the end comes to obtaining fixed points which are defined by
simultaneous zero of the both $\beta$-functions. Among all the possible
fixed points
one is interested only in those in the ranges $u^*>0,v^*\leq0$ and only in
stable ones where the stability means that two eigenvalues $b_1,b_2$ of
the stability matrix
$B= \partial \beta_{u_i}/ \partial u_j|_{u^*_i}$, $u_i\equiv\{u,v\}$
are positive or possess positive real
parts. The structure of the $\beta$-functions
(\ref{beta_u})-(\ref{beta_v}) yields the possibility of four solutions
for the fixed points. The first two $\{ u^*=0, v^*=0 \}$ and $\{
u^*=0, v^*>0 \}$ in our case at $d<4$ are out of physical interest,
while the second pair which consists of pure $\{ u^*>0, v^*=0 \}$ and
mixed $\{ u^*>0, v^*<0 \}$ points, are responsible for two possible
critical regimes. The critical behaviour of the diluted model
coincides with that of the pure model when the pure fixed point
appears to be stable. If the mixed point is stable, the {\it new}
(diluted) critical behaviour of the system takes place.  The type of
the critical behaviour depends on the number $m$ of the
order parameter components and on the dimensionality $d$: at any $d, 2
\leq d < 4$ a system with large enough $m$ is not sensitive to the
weak dilution in the sense that asymptotic values of critical exponents
do not change; only starting from some marginal value
$m_c$, at $m<m_c$ a mixed fixed point becomes stable and the crossover
to the random critical behaviour occurs. The problem of determining
$m_c$ as a function of $d$ will be discussed later. Now we would like
to state that $m_c \geq 1$ for any $d, 2 \leq d < 4$, and thus just
the mixed fixed point governs the asymptotic critical behaviour of the
diluted Ising model.

If one attempts to find the fixed
points from the $\beta$-functions (\ref{beta_u})-(\ref{beta_v})
without resummation, there always appears only the Gaussian
$\{ u^*=0, v^*=0 \}$ trivial
solution; the existence of the rest possible three fixed points
depends on the concrete details of the $\beta$-functions portions in
the braces in expressions (\ref{beta_u})-(\ref{beta_v}).
In a $3d$ case it appears that without a resummation the non-trivial mixed
fixed point does not exist in one-, two- and four-loop approximations
\cite{MayerSokolovShalaev'89,Mayer'89}. It
is only the three-loop approximation where all the four solutions of the set
of equations (\ref{fixedpoint}) exist \cite{SokolovShalaev'81}. In
figure \ref{fig1} we show the behaviour of the non-resummed $\beta$-functions
of the three-dimensional weakly diluted Ising model in the three-loop
approximation. Resummed functions are shown in the same approximation
in figure \ref{fig2}.
Note that in this approximation
the shape of the functions remains alike in the
region of small couplings $u$ and $v$.
Fixed points correspond to the crossing of
the lines $\beta_u=0, \beta_v=0$ as it is demonstrated in figures
\ref{fig3}, \ref{fig4}.
The left-hand column in figures  \ref{fig3}, \ref{fig4} shows the lines of
zeros of non-resummed $\beta$-functions in three-dimensions in one-, two-,
three- and four-loop (results of
\cite{MayerSokolovShalaev'89,Mayer'89}) approximations.
One can see in the figures that without resummation all
non-trivial solutions are obtained only within the three-loop level of
the perturbation theory. In the next order all fixed points disappear
which is a strong evidence of their accidental origin. At any
arbitrary $d$, $2 \leq d <4$ the qualitative behaviour of the
functions is very similar to that shown in figures \ref{fig3} and \ref{fig4}.

As it has already been mentioned, in order to reestablish the lost pure and
mixed points one applies the resummation procedure to $\beta$-functions.
In the three-dimensional space the result of resummation is
illustrated by the right-hand column in figures  \ref{fig3} and \ref{fig4}.
Here we have used the Chisholm-Borel resummation technique choosing Chisholm
approximant in the form discussed in the previous Section
with $p=1$ in successive approximation in the number
of loops. The icons in the figures which correspond to a one-loop level
are the visual proof of the degeneracy of the $\beta$-functions in
this order of the perturbation theory: the plots of root-lines are
parallel independently of resummation. The rest three images in the
right-hand columns are a good graphic demonstration of the reliability of
the Chisholm-Borel resummational method: two-, three- and four-loop
pictures are quantitatively similar, the coordinates of the pure and
mixed point are close.

The numerical results of our study are given in table \ref{table1}.
Here, the
coordinates of the stable mixed fixed point and the values of the critical
exponents of the quenched weakly diluted Ising model are listed as
functions of $d$ between $d=2$ and $d=3.8$. The eigenvalues $b_1$ and
$b_2$ of the stability matrix are given as well.

It was already noted that the values of $\gamma$-functions in a stable point
yield the numerical characteristics of the critical behaviour of the model.
For example, given the resummed functions $\gamma^{Res}_{\phi}$ and
$\bar{\gamma}^{Res}_{\phi^2}$, the pair of equations
\begin{eqnarray}
\gamma^{Res}_{\phi}(u^*,v^*) &=& \eta, \\
\bar{\gamma}^{Res}_{\phi^2}(u^*,v^*) &=& 2 - \nu^{-1} - \eta
\end{eqnarray}
allows us to find the exponents $\eta$ and $\nu$. All other exponents can be
obtained from the familiar scaling laws.

However, one can proceed in a
different way. That is, by means of the scaling laws it is
possible to reconstitute the expansion in coupling constants
of any exponent of interest or of any combination of exponents, and only
after that to apply the resummation procedure. If exact calculation were
performed the answer would not depend on the sequence of
operations. However, this is not the case for the present approximate
calculations. We have chosen the scheme of computing where the
resummation procedure was applied to the combination $\nu^{-1}-1 =
1-\bar{\gamma}_{\phi^2}-\gamma_{\phi}$ and $\gamma^{-1} =
(2-\bar{\gamma}_{\phi^2}-\gamma_{\phi})/(2-\gamma_{\phi})$. The exponents
$\alpha$, $\beta$ and $\eta$ have been calculated on the basis of numerical
values of the exponents $\gamma$ and $\nu$. The resummation
scheme appears to be quite insensitive to the choice of the parameter $p$
given by (\ref{Leroy}), (\ref{p}).
However note, that computations have been performed here, as
well as in \cite{HolovatchShpot'92}, with $p=1$.

Comparing our data from table \ref{table1} for the critical exponents at
$d=2$ with the results for the pure Ising model one can see that the exponent
$\gamma$ differs from the exact value $7/4$ by the order of $5\%$, the
exponent $\nu$ is smaller from the exact value $\nu=1$ less than by $4\%$.
This confirm the conjecture that the critical behaviour of the weakly diluted
quenched Ising model at $d=2$ within logarithmic correction coincide with
that of the pure model (see \cite{Shalaev'94} for review).
It is also interesting
to compare numbers given in table \ref{table1} with those obtained for
general $d$ within the 2-loop approximation \cite{HolovatchShpot'92}: all the
exponents of the three-loop level lie slightly farther from the expected
exact values of Onsager than those of the two-loop approximation. This may be
explained by the oscillatory nature of approaching to the exact values
depending on the order of the perturbation theory. It is also interesting to
note that the two-loop approximation yields better estimates for the heat
capacity critical exponent $\alpha$ for all $d$ in the range under
consideration. Namely, in accordance with the Harris criterion, the
exponent $\alpha$ for the diluted Ising system should remain negative. This
picture is confirmed much better by the two-loop approximation where $\alpha$
is negative in the whole range of $d$, unlike the three-loop level of the
perturbation theory, the results of which yield $\alpha>0$ for $2 \leq d \leq
2.8$.

However, table \ref{table1} shows that the next (third) order does improve our
understanding of the critical behaviour of the model in general
dimensions. The results of the two-loop calculations
\cite{HolovatchShpot'92} show that starting from some marginal space
dimension the approach to the stable point becomes oscillatory: the
eigenvalues $b_1$ and $b_2$ turn to be complex possessing positive
real parts.
This is an artifact of the calculation scheme and therefore it
was expected \cite{HolovatchShpot'92}
that by increasing the accuracy of calculations one decreases
the region of $d$ which corresponds to the complex eigenvalues. It is
really the case. In the three-loop approximation the region of complex $b_1,
b_2$ is bounded from below by $d=3.3$, whereas in the two-loop approximation
\cite{HolovatchShpot'92} the corresponding value is lower and is equal
to $d=2.9$. Thus, the region of $d$ characterized by the oscillatory
approach to the stable fixed point shrinks with the increase of the
order of the perturbation theory.

The comparison of the three-dimensional value of $\nu$ with the four-loop
result \cite{Mayer'89} $\nu=0.6701$ gives the accuracy of $0.05 \%$
for our computations (compare with $1\%$ for two-loops, where the value
$\nu(d=3)=0.678$ was obtained).  Thus, it may be stated that the general
accuracy of calculations decreases when passing from $d=4$ to $d=2$ which,
in particular, results from the fact that our approach is asymptotically
exact at upper critical dimension $d=4$.

The comparison of the present results with the other data available is
provided by figure \ref{fig5}.  Here, the behaviour of the correlation length
critical exponent $\nu$ obtained by different methods is demonstrated in
general dimensions. The results of the massive field-theoretical scheme are
plotted by solid (three-loop approximation; the present paper) and dashed
(two-loop approximation; ref. \cite{HolovatchShpot'92}) lines. One can see
that the two lines practically coincide far enough from $d=2$, in particular,
both lie very close to the most accurate result for $d=3$ \cite{Mayer'89}
which is shown by the box. The application of the scaling-field method
\cite{NewmanRiedel'82}
yields numbers shown in
figure \ref{fig5} by stars. The limit from below ($d=2.8$) of the
method applicability is caused by the truncation of the set of
scaling-field equations, which was considered in
\cite{NewmanRiedel'82}.
One can also attempt to obtain some results by resumming the
$\sqrt{\varepsilon}$-ex\-pan\-si\-on which is known for the diluted
Ising model up to three-loop order \cite{Shalaev'77,JayaprakashKatz'77}
and for the exponents $\nu$ and $\eta$ reads:
\begin{eqnarray}
\nu &=& \frac{1}{2} + \frac{1}{4}\Big( \frac{6}{53}\varepsilon \Big)^{1/2} +
\frac{535 - 756\zeta(3)}{8(53)^2}\varepsilon, \label{nu_sqrteps} \\
\eta &=& - \frac{\varepsilon}{106} +
\Big( \frac{6}{53} \Big)^{1/2}\frac{9}{(53)^2}
(24 + 7\zeta(3))\varepsilon^{3/2},
\end{eqnarray}
where $\zeta(3)\approx 1.202$ is Rieman's zeta function.
The corresponding results are shown by
open diamonds. They were obtained by applying the Pad\'e-Borel resummation
scheme to the series of $\sqrt{\varepsilon}$- expansion (\ref{nu_sqrteps})
\cite{Shalaev'77,JayaprakashKatz'77}.  The value of $\nu$ obtained in such a
way is of physical interest only very close to $d=4$. Even in the next orders
of the expansions the values of critical exponents are not improved
\cite{HolovatchYavors'kii'97}; this is an evidence of the
$\sqrt{\varepsilon}$-expansion unreliability in tasks like the one under
consideration. To compare, one can state that the situation with the applied
in the present paper theoretical scheme is contrary to the
$\sqrt{\varepsilon}$-expansion. While the two-loop approximation is valid in
ranges $2 \leq d < 3.4$, the next order of the perturbation theory enlarges
the upper bound up to $d=3.8$. One can expect that the next steps within the
perturbation theory will allow one to obtain the description of the critical
behaviour of the model with enough accuracy for any $d$, $2 \leq d <4$.

Let us recall now that expressions (\ref{beta_u})-(\ref{gamma}) for the
$RG$-functions, as well as their three-loop parts listed in the Appendix,
allow us to study asymptotic critical properties of the $mn$-vector model
with arbitrary $m$ and $n$ in arbitrary $d$ not only for the case
$m=1$, $n=0$. In particular, by keeping $m$ as an arbitrary number and
putting $n=0$ one can obtain the numerical estimates for the marginal
order parameter component number $m_c$ which divides the diluted
(governed by the mixed fixed point) asymptotic critical behaviour from
the pure one, when the $O(m)$-symmetric fixed point remains stable. In
accordance with the Harris criterion the case $m=m_c$ corresponds to
zero of the heat capacity critical exponent $\alpha$ of the model. One
may extract the value of $m_c$ from this condition. However, the above
discussed results of the three-loop approximation do not yield enough
accuracy for $\alpha$. Alternatively, the fixed mixed point should coincide
with the pure fixed
point at $m=m_c$, which in particular means that
$v^*(m=m_c)|_{mixed}=0$. The last condition was chosen as a basis of our
calculation.  The appropriate numbers of the present three-loop approximation
(thick solid line) together with the data of the two-loop approximation
(dashed line) \cite{HolovatchShpot'92} are shown in figure \ref{fig6}. The
result of $\varepsilon$-expansion $m_c=4-4 \varepsilon$ is depicted by the
thin solid line. In the three-loop approximation we obtain $m_c=1.40, d=2$
and $m_c=2.12, d=3$. These values are to be compared with the exact results
of Onsager which yield $m_c=1$ at $d=2$, and the theoretical estimate
$m_c=1.945 \pm 0.002$ \cite{Bervillier'86}. One can see that the two-loop
results are closer to the expected values for both $d=2$ and $d=3$. For a
two-dimensional case the two-loop value $m_c=1.19$ \cite{HolovatchShpot'92}
differs from the exact one by $20 \%$, while the three-loop number decreases
the accuracy to $40 \%$. The case $d=3, m_c>2$ contradicts the suggestion
that the $xy$-model asymptotic critical behaviour should not change under
dilution in three-dimensions. The reason for decreasing the calculation
accuracy with increasing the order of the perturbation theory may lie in
oscillatory approach to the exact result. One can expect that already
the four-loop case will improve the estimates for $m_c$ for all $2 \leq d
<4$. Let us also note that the determination of $m_c$ may serve as a test for
improving the resummation scheme.

%%%%%%%%%%%%%%%%%%%%%%%%%%%%%%%%%%%%%%%%%%%%%%%%%%%%%%%%%%%%%%%%%%%%%%%%%%%
%                           Section 3.
%                          Conclusions
%%%%%%%%%%%%%%%%%%%%%%%%%%%%%%%%%%%%%%%%%%%%%%%%%%%%%%%%%%%%%%%%%%%%%%%%%%%

\section{Conclusions}

The goal of this paper is to study the critical behaviour of the weakly
diluted quenched Ising model in the case when the space dimension $d$
continuously changes from $d=2$ to $d=4$.

As it was mentioned in the Introduction, the study of the pure Ising
model at arbitrary $d$, which corresponds to a scalar
field-theoretical model with one coupling constant, is the subject of a
great deal of papers. It is not the case for the model with a more
complicated symmetry. In particular, here we study a model with two
couplings corresponding to terms of different symmetry in the
Lagrangian (\ref{Lagrangian}). Such a problem was
studied previously on the basis of the scaling-field method
\cite{NewmanRiedel'82}, and field-theoretical fixed dimension
renormalization group calculations within a two-loop level of
the perturbation theory are available \cite{HolovatchShpot'92}.

Our calculations hold within the theoretical scheme of
\cite{HolovatchKrokhmal's'kii'94,HolovatchShpot'92}. This approach appears to
be one amidst other possible calculation schemes for many tasks; however, in
our case it seems to have no alternatives within the field-theoretical
approach.

Being asymptotic, the resulting series for the $RG$-functions are to be
resummed. In the present study we have chosen the Pad\'e-Borel and the
Chisholm-Borel resummation techniques. Restricting ourselves to
analytic expressions for the resummed functions, we present numerical
data mainly obtained on the basis of the Chisholm-Borel resummation
technique. Note that the absence of any information on the high-order
behaviour of the obtained series for the $RG$-functions does not allow
one to apply other resummation schemes, e.g. those based on the conformal
mapping technique \cite{LeGuillouZinnJustin'80}.

The quantitative description of the critical behaviour of the model is
steady from the point of view of passing from the two- to the three-loop
approximation.  Smaller agreement between the two- and the three-loop
approximations at $d$ far away from $d=4$ may be explained in a way
that the precision of computing falls down with the increase of the
expansion parameter which takes place at decrease of $d$.
The real parts of eigenvalues corresponding to the mixed point seem
to remain positive up to $d=4$, which testifies that at
arbitrary $d$ the weakly diluted quenched Ising model is described by the
mixed fixed point.

\vspace{2ex}
The work was supported in part by the Ukrainian Foundation of Fundamental
Studies (grant No 24/173).

%%%%%%%%%%%%%%%%%%%%%%%%%%%%%%%%%%%%%%%%%%%%%%%%%%%%%%%%%%%%%%%%%%%%%%%%%%%
%                           The bibliography
%%%%%%%%%%%%%%%%%%%%%%%%%%%%%%%%%%%%%%%%%%%%%%%%%%%%%%%%%%%%%%%%%%%%%%%%%%%

%%%%%%%%%%%%%%%%%%%%%%%%%%%%%%%%%%%%%%%%%%%%%%%%%%%%%%%%%%%%%%%%%%%%%%%%%%%
%                           The Appendix
%%%%%%%%%%%%%%%%%%%%%%%%%%%%%%%%%%%%%%%%%%%%%%%%%%%%%%%%%%%%%%%%%%%%%%%%%%%
\newpage
\section*{Appendix}

Here we have collected the most lengthy expressions for the three-loop
contributions to the $RG$-functions.
The three-loop part of the $\beta_u$-function reads:
\begin{eqnarray}
\beta_u^{3LA}(u,v) = \beta_u^{3,0} u^3 + \beta_u^{2,1} u^2 v +
\beta_u^{1,2} u v^2 + \beta_u^{0,3} v^3,
\end{eqnarray}
where

\begin{eqnarray}
\beta_u^{3,0} & = & -\frac{1}{(m+8)^3} \Big [ -4(31m^2+430m+1240)i_1 +
(m+8)(m+2) \times
\nonumber\\&&
\Big (-(3d+8)i_2+12(i_3+i_8) \Big )+
48(m^2+20m+60)i_4 +
\nonumber\\&&
24(2m^2+21m+58)i_5+6(3m^2+22m+56)i_6 +
\nonumber\\&&
24(5m+22)i_7 +
8(4m^2+61m+178) \Big ];
\nonumber\\
\beta_u^{2,1} & = & -\frac{2}{(m+8)^2(mn+8)} \Big [ -12(17m^2+256m+780)i_1 +
\nonumber\\&&
(m+2)\Big(-(3dm+42d+16m+80)i_2 +
12(m+14)i_3 +
\nonumber\\&&
18(m+8)i_8 \Big)+
24(3m^2+70m+224)i_4 +
\nonumber\\&&
6(15m^2+158m+448)i_5+
6(3m^2+32m+100)i_6 +
\nonumber\\&&
48(5m+22)i_7+6(9m^2+146m+448) \Big ];
\nonumber\\
\beta_u^{1,2} & = & -\frac{1}{(m+8)(mn+8)^2}
\Big [ -12(19m^2n+80mn+470m+2032)i_1 -
\nonumber\\&&
\Big ( 8(mn+8)(3d+4) +
m(3dmn+40mn+78d+176) \Big )i_2 +
\nonumber\\&&
12(m^2n+8mn+26m+64)i_3 +
48(m^2n+8mn+68m +
\nonumber\\&&
292)i_4+12(11m^2n+34mn+136m+584)i_5 +
\nonumber\\&&
6(m^2n+8mn+50m+256)i_6+576(m+5)i_7+
\nonumber\\&&
36(m+2)(mn+8)i_8 +
12(5m^2n+22mn+136m+584) \Big ];
\nonumber\\
\beta_u^{0,3} & = & -\frac{4}{(mn+8)^3}
\Big [ 6(mn+10)(-2(mn+23)i_1+3i_6)+
\nonumber\\&&
(mn+2) \Big(-(4mn+9d+8)i_2+36i_3+3(mn+8)i_8\Big)+
\nonumber\\&&
72(3mn+22)i_4+9(m^2n^2+14mn+88)i_5+
\nonumber\\&&
24(mn+14)i_7+3(m^2n^2+38mn+264) \Big ].
\nonumber
\end{eqnarray}

The three-loop part of the $\beta_v$-function reads:
\begin{eqnarray}
\beta_v^{3LA}(u,v) = \beta_v^{0,3} v^3 + \beta_v^{1,2} u v^2 +
\beta_v^{2,1} u^2 v + \beta_v^{3,0} u^3,
\end{eqnarray}
where

\begin{eqnarray}
\beta_v^{0,3} & = & -\frac{1}{(mn+8)^3}
\Big [ -4(31m^2n^2+430mn+1240)i_1+
(mn+8) \times
\nonumber\\&&
(mn+2) \Big(-(3d+8)i_2+12(i_3+i_8)\Big)+
48(m^2n^2+
\nonumber\\&&
20mn+60)i_4+24(2m^2n^2+21mn+58)i_5+6(3m^2n^2+
\nonumber\\&&
22mn+56)i_6+24(5mn+22)i_7+8(4m^2n^2+61mn+178) \Big ];
\nonumber\\
\beta_v^{1,2} & = & -\frac{4(m+2)}{(mn+8)^2(m+8)}
\Big [ -4(28mn+275)i_1 -
\nonumber\\&&
(3dmn+4mn+15d+56)i_2+12(mn+5)i_3 +
\nonumber\\&&
24(2mn+27)i_4+3(13mn+100)i_5+6(3mn+13)i_6 +
\nonumber\\&&
96i_7+9(mn+8)i_8+(29mn+316) \Big ];
\nonumber\\
\beta_v^{2,1} & = & -\frac{m+2}{(mn+8)(m+8)^2}
\Big [ -12(mn+42m+224)i_1-
\nonumber\\&&
(3dmn+12dm-8mn+48d+16m+256)i_2 +
\nonumber\\&&
12(mn+4m+16)i_3+48(5m+34)i_4 +
\nonumber\\&&
12(13m+56)i_5+6(3mn+14m+40)i_6+
\nonumber\\&&
144i_7+36(m+8)i_8+12(11m+64) \Big ];
\nonumber\\
\beta_v^{3,0} & = & -\frac{2(m+2)}{(m+8)^3}
\Big [ -4(11m+70)i_1-3(dm+2d+16)i_2+
\nonumber\\&&
6(m+2)(2i_3+3i_6)+
2(m+8)(12i_4+3i_5+3i_8+5) \Big ].
\nonumber
\end{eqnarray}

The three-loop part of the $\gamma_{\phi}$-function reads:
\begin{eqnarray}
\gamma_{\phi}^{3LA}(u,v) & = &
- \Big [
\frac{m+2}{(m+8)^2} u^3 + \frac{3(m+2)}{(m+8)(mn+8)} u^2 v +
\\&&
\frac{3(m+2)}{(m+8)(mn+8)} u v^2 + \frac{mn+2}{(mn+8)^2} v^3
\Big ](3i_8-4i_2).
\nonumber
\end{eqnarray}
The three-loop part of the $\bar{\gamma}_{\phi^2}$-function reads:
\begin{eqnarray}
\bar{\gamma}_{\phi^2}(u,v) =
\bar{\gamma}_{\phi^2}^{3,0} u^3 + \bar{\gamma}_{\phi^2}^{2,1} u^2 v +
\bar{\gamma}_{\phi^2}^{1,2} u v^2 + \bar{\gamma}_{\phi^2}^{0,3} v^3,
\end{eqnarray}
where
\begin{eqnarray}
\bar{\gamma}_{\phi^2}^{3,0} & = & \frac{m+2}{(m+8)^3}
\Big [ -4(11m+70)i_1+(m+2)\Big(-(3d-8)i_2+
\nonumber\\&&
12i_3+18i_6\Big)+2(m+8)(12i_4+3i_5+5) \Big ];
\nonumber\\
\bar{\gamma}_{\phi^2}^{2,1} & = & \frac{m+2}{(m+8)^2(mn+8)}
\Big [ -12(mn+10m+70)i_1 +
\nonumber\\&&
(mn+2m+6)\Big(-(3d-8)i_2+12i_3+18i_6\Big) +
\nonumber\\&&
6(m+8)(12i_4+3i_5+5) \Big ];
\nonumber\\
\bar{\gamma}_{\phi^2}^{1,2} & = & \frac{3(m+2)}{(m+8)(mn+8)^2}
\Big [ -4(11mn+70)i_1 +
\nonumber\\&&
(mn+2)\Big(-(3d-8)i_2+12i_3+18i_6\Big) +
\nonumber\\&&
2(mn+8)(12i_4+3i_5+5) \Big ];
\nonumber\\
\bar{\gamma}_{\phi^2}^{0,3} & = & \frac{mn+2}{(mn+8)^3}
\Big [ -4(11mn+70)i_1 +
\nonumber\\&&
(mn+2)\Big(-(3d-8)i_2+12i_3+18i_6\Big) +
\nonumber\\&&
2(mn+8)(12i_4+3i_5+5) \Big ].
\nonumber
\end{eqnarray}

\label{pend}

%%%%%%%%%%%%%%%%%%%%%%%%%%%%%%%%%%%%%%%%%%%%%%%%%%%%%%%%%%%%%%%%%%%%%%%%%%%
%                           Figure captions
%%%%%%%%%%%%%%%%%%%%%%%%%%%%%%%%%%%%%%%%%%%%%%%%%%%%%%%%%%%%%%%%%%%%%%%%%%%
\newpage
\centerline{FIGURE CAPTIONS.}

\begin{figure}[htbp]
\begin{centering}
\setlength{\unitlength}{1mm}
\begin{picture}(126,90)
\epsfxsize=126mm
\epsfysize=90mm
\put(0,0){\epsffile[29 29 816 584]{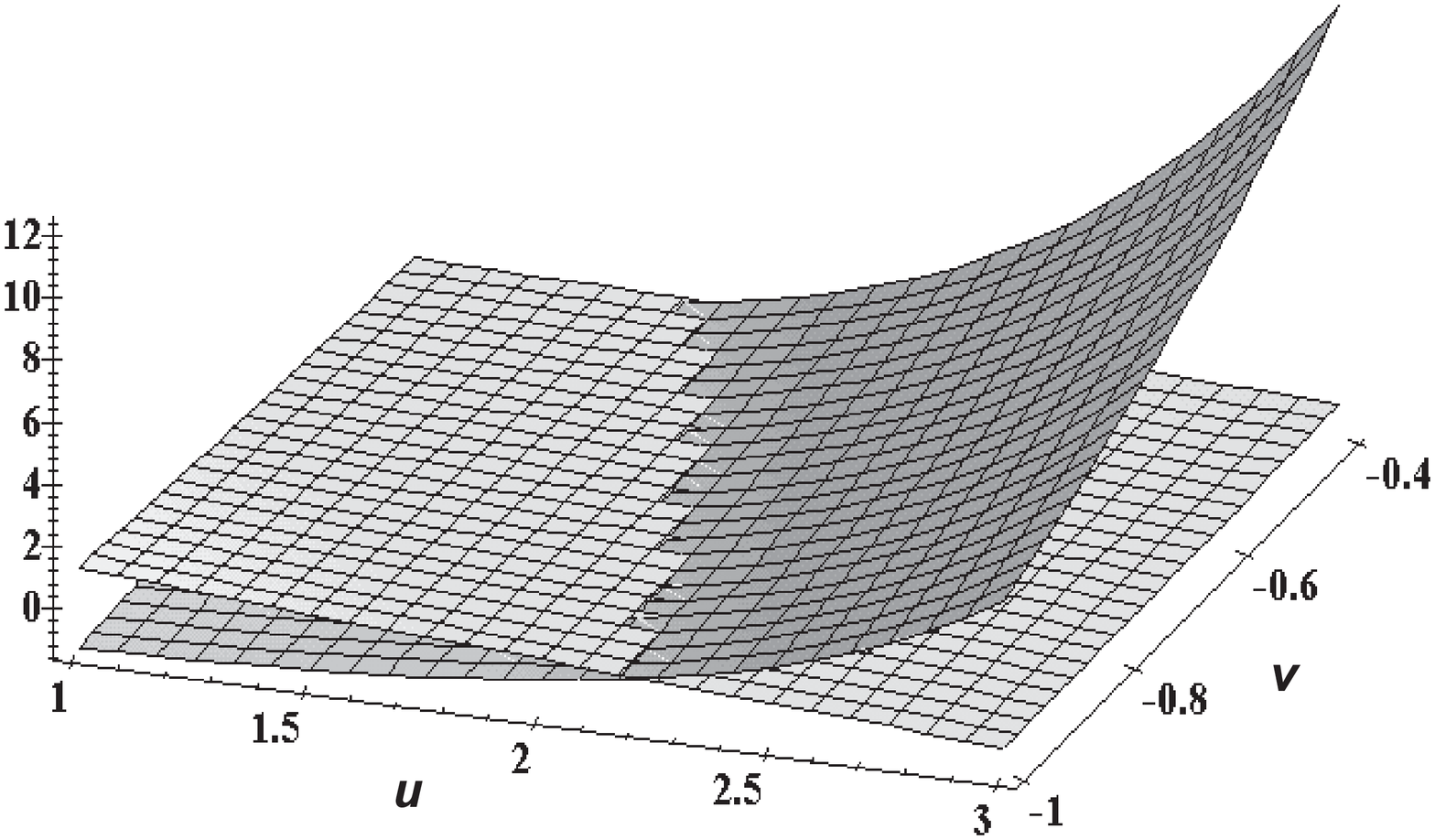}}
\end{picture}\\
\end{centering}
\caption{\label{fig1}
}
\end{figure}

\noindent
Figure \ref{fig1}
{The non-resummed $\beta$-functions in the three-loop
approximation; $d=3, m=1, n=0$. The dark surface corresponds to the
$\beta_u$-function. }
\vspace{2ex}

\begin{figure}[htbp]
\begin{centering}
\setlength{\unitlength}{1mm}
\begin{picture}(126,90)
\epsfxsize=126mm
\epsfysize=90mm
\put(0,0){\epsffile[22 20 822 582]{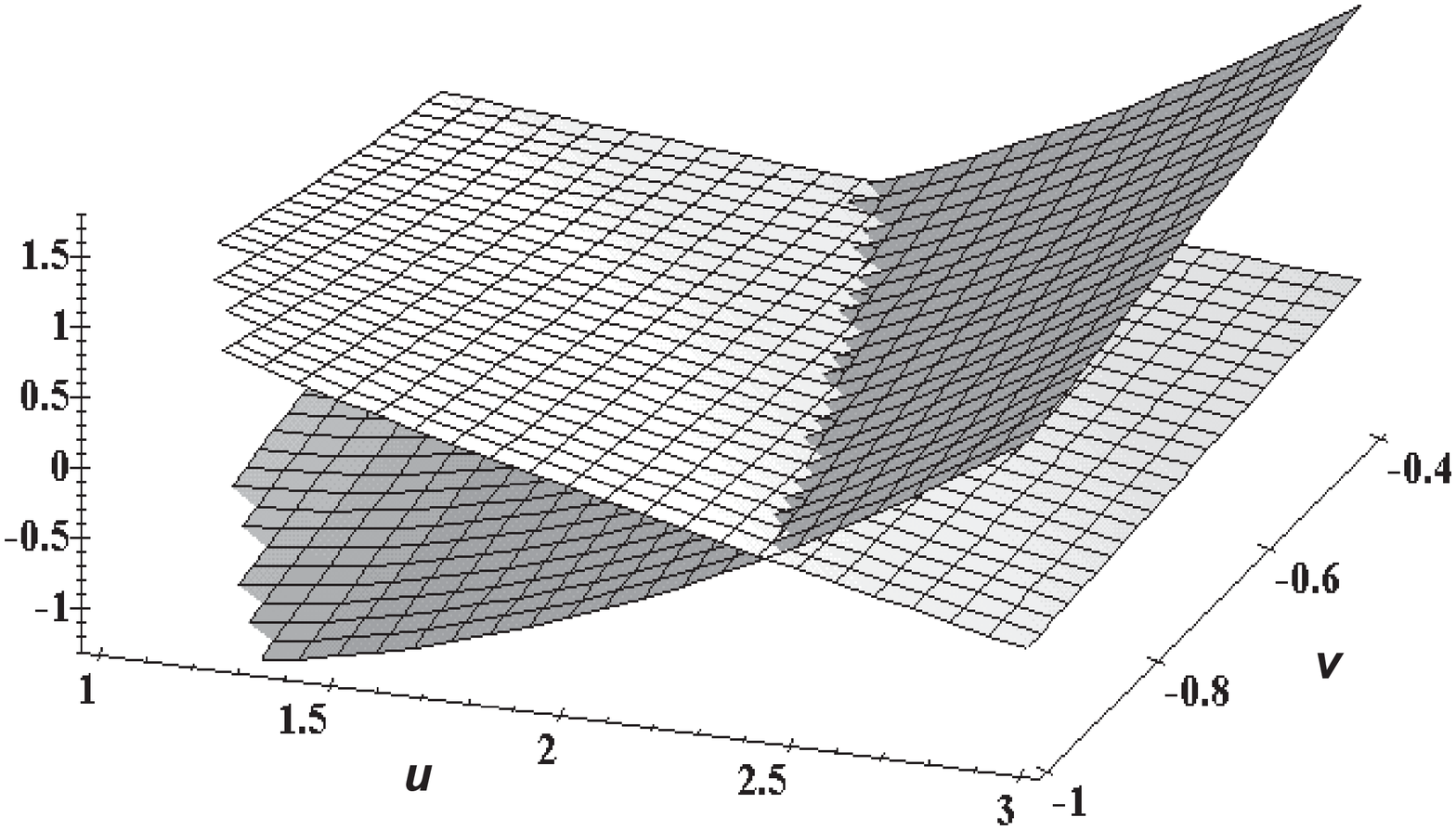}}
\end{picture}\\
\end{centering}
\caption{\label{fig2}
}
\end{figure}

\noindent
Figure \ref{fig2}
{The Chisholm-Borel resummed
$\beta$-functions in the three-loop approximation; $d=3, m=1, n=0$.
The dark surface corresponds to the
$\beta_u$-function. }
\vspace{2ex}

\begin{figure}[htbp]
\begin{centering}
\setlength{\unitlength}{1mm}
\begin{picture}(126,64)
\epsfxsize=126mm
\epsfysize=64mm
\put(0,0){\epsffile[7 8 832 584]{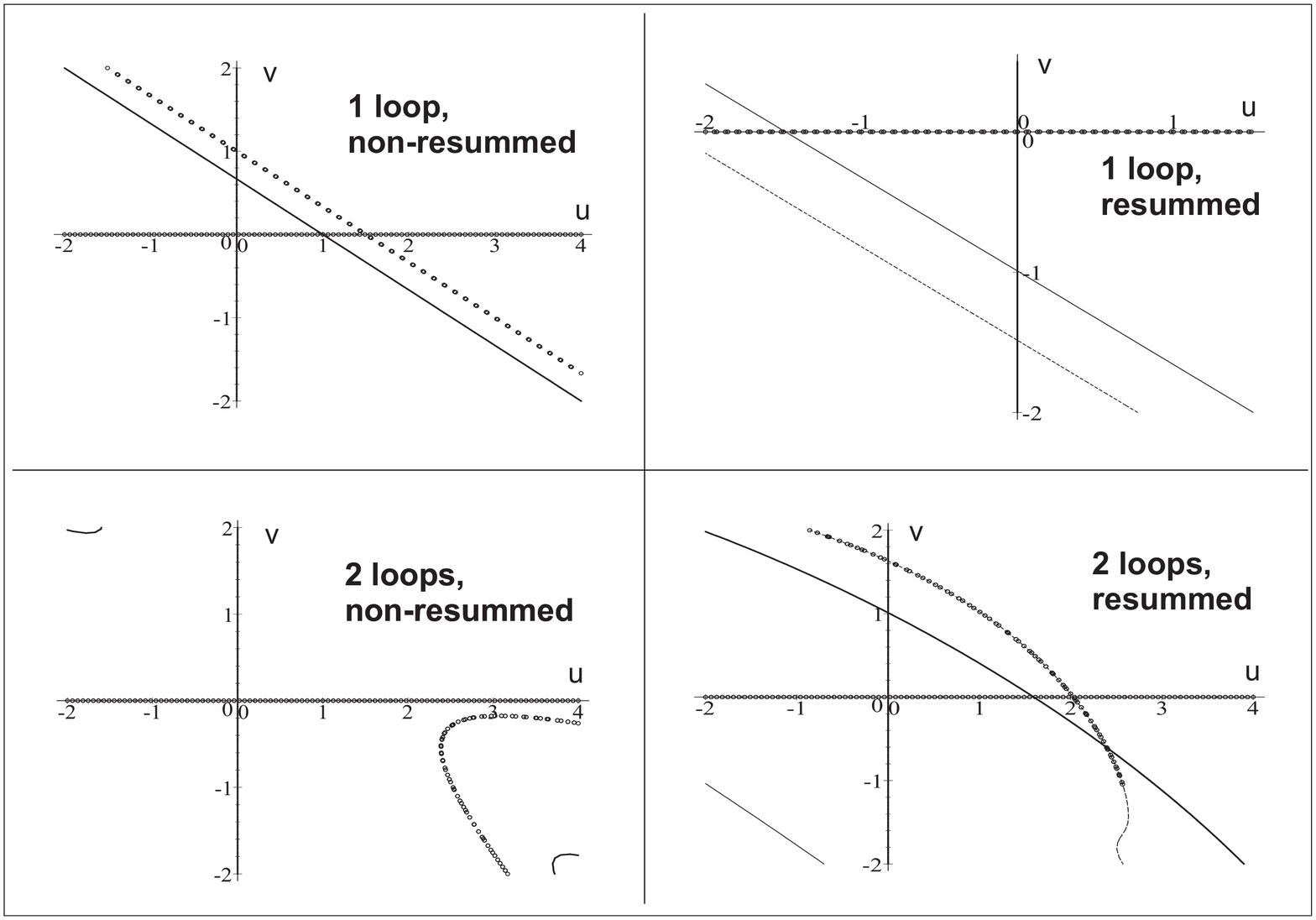}}
\end{picture}\\
\end{centering}
\caption{\label{fig3}
}
\end{figure}

\noindent
Figure \ref{fig3}
{The lines of zeros of non-resummed (left-hand column) and resummed
by the Chi\-sholm-Borel method (right-hand column) $\beta$-functions for
$m=1, n=0$ in different orders of the perturbation theory:
one- and two-loop approximations. Circles correspond to $\beta_u=0$,
thick lines depict $\beta_v=0$. Thin solid and dashed lines show the
roots of the analytically continued functions $\beta_u$ and $\beta_v$
respectively. One can see the appearance of the mixed fixed point
$u>0, v<0$ in the two-loop approximation for the resummed
$\beta$-functions.}
\vspace{2ex}

\begin{figure}[htbp]
\begin{centering}
\setlength{\unitlength}{1mm}
\begin{picture}(126,64)
\epsfxsize=126mm
\epsfysize=64mm
\put(0,0){\epsffile[7 8 832 584]{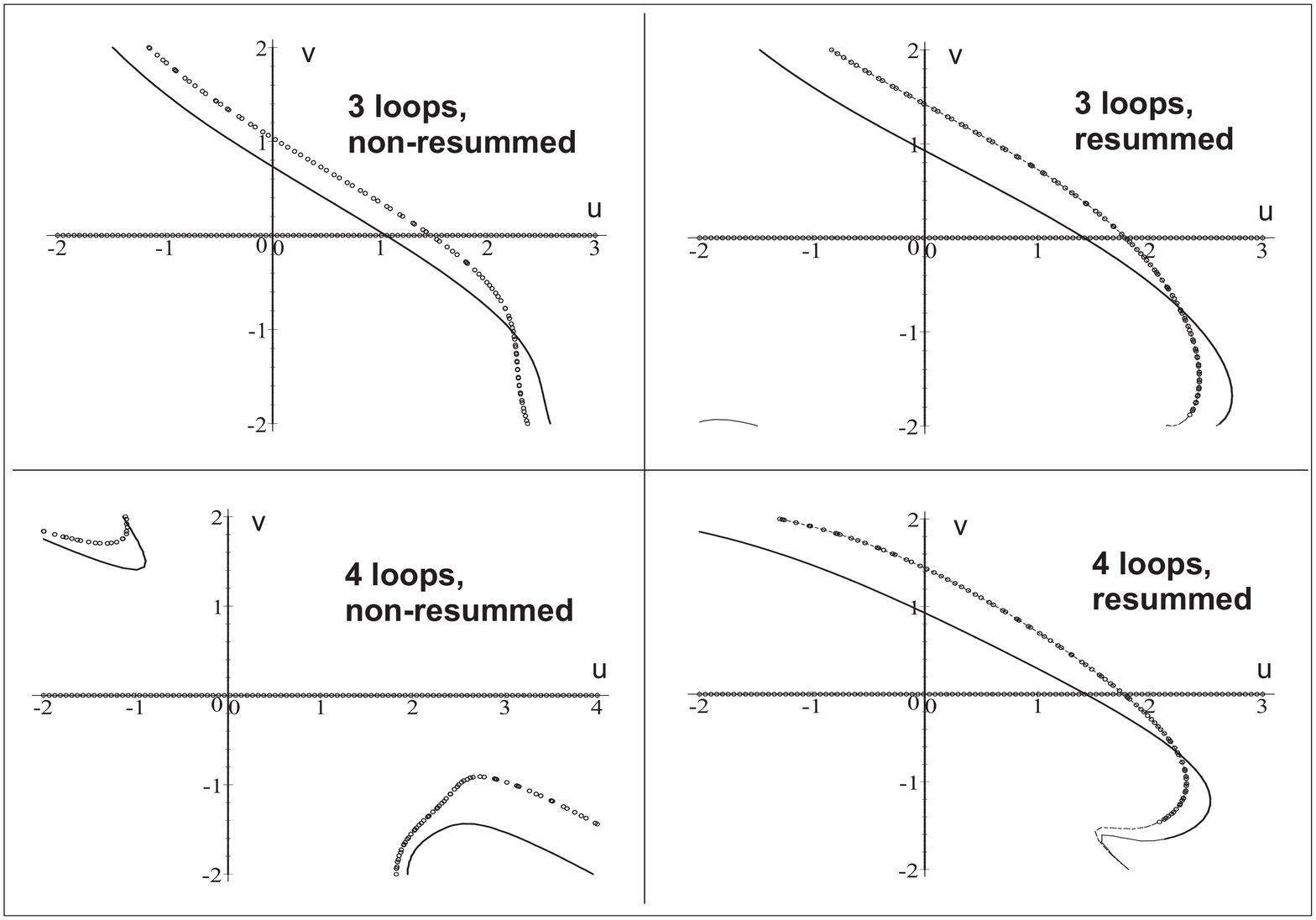}}
\end{picture}\\
\end{centering}
\caption{\label{fig4}
}
\end{figure}

\noindent
Figure \ref{fig4}
{The lines of zeros of non-resummed (left-hand column) and resummed
by the Chi\-sholm-Borel method (right-hand column) $\beta$-functions for
$m=1, n=0$ in three- and four-loop approximations. The notations
are the same as in figure \ref{fig3}. Close to the mixed fixed point
the behaviour of the resummed functions
remains alike with the increase of the order of approximation. This
is not the case for non-resummed functions. }
\vspace{2ex}

\noindent
Figure \ref{fig5}.
{The correlation length critical exponent
$\nu$ of the weakly diluted Ising model as a function of the
space dimension $d$. The results of two-
\protect\cite{HolovatchShpot'92}
and three-loop (the present paper) approximations are shown by
the dashed and the solid lines respectively, the square reflects the number
of the  four-loop approximation
\protect\cite{Mayer'89}
at $d=3$, stars correspond to work
\protect\cite{NewmanRiedel'82}
and open diamonds refer to the resummed
$\protect\sqrt\varepsilon$-expansion. }
\vspace{2ex}

\noindent
Figure \ref{fig6}.
{The dependence of the marginal order parameter component
number $m_c$ on the space dimension $d$. Two- and three-loop results
are shown by the dashed and thick solid lines respectively, the
$\varepsilon$-expansion data $m_c=4-4 \varepsilon$ are depicted by the
thin solid line.}

%%%%%%%%%%%%%%%%%%%%%%%%%%%%%%%%%%%%%%%%%%%%%%%%%%%%%%%%%%%%%%%%%%%%%%%%%%%
%                           Figures
%%%%%%%%%%%%%%%%%%%%%%%%%%%%%%%%%%%%%%%%%%%%%%%%%%%%%%%%%%%%%%%%%%%%%%%%%%%
\newpage

\begin{figure}[htbp]
\begin{center}
\input{nudil.pic}
\end{center}
\caption{\label{fig5}
}
\end{figure}

\begin{figure}[htbp]
\begin{centering}
\setlength{\unitlength}{1mm}
\begin{picture}(126,64)
\epsfxsize=126mm
\epsfysize=64mm
\put(0,0){\epsffile[33 32 811 561]{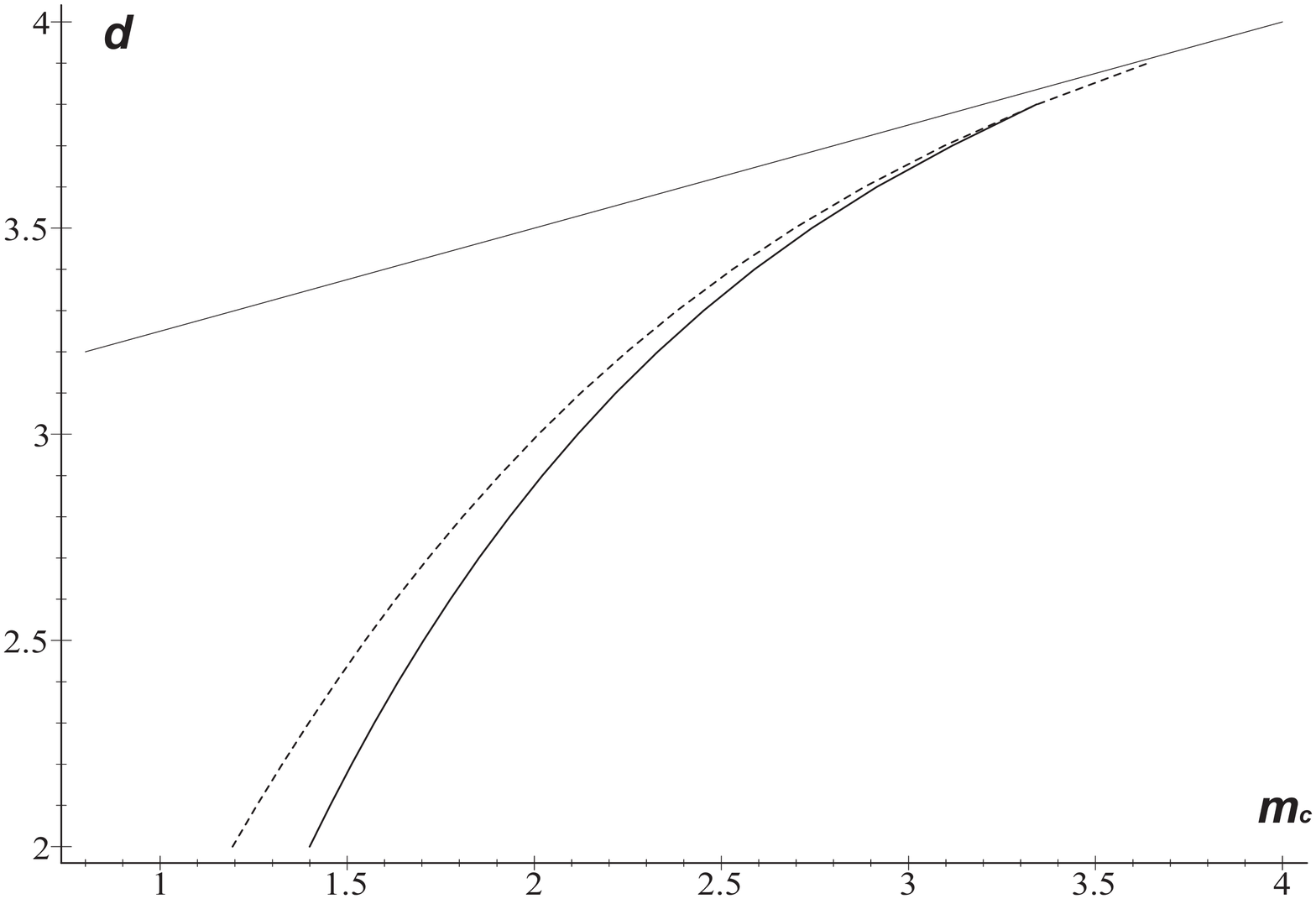}}
\end{picture}\\
\end{centering}
\caption{\label{fig6}
}

\end{figure}

%%%%%%%%%%%%%%%%%%%%%%%%%%%%%%%%%%%%%%%%%%%%%%%%%%%%%%%%%%%%%%%%%%%%%%%%%%%
%                           Tables
%%%%%%%%%%%%%%%%%%%%%%%%%%%%%%%%%%%%%%%%%%%%%%%%%%%%%%%%%%%%%%%%%%%%%%%%%%%
\newpage

\begin{table}
\caption{\label{table1}
{The stable point coordinates, critical exponents and
the eigenvalues of the stability matrix of the weakly diluted Ising
model at arbitrary $d$. The three-loop approximation (the superscript ''$c$''
denotes that real parts of the corresponding eigenvalues are
given). }
}
\begin{center}
\tabcolsep1.6mm
\begin{tabular}{lllllllll}
\hline
\hline
$d$ & $u^*$ & $v^*$ & $\gamma$ & $\nu$ & $\alpha$ & $\eta$ & $b_1$ &
$b_2$\\
\hline \\
2.0 & 2.0268 &-0.2802&1.840&0.966& 0.067&0.097 &0.2176 &1.5189\\
2.1 & 2.0327 &-0.3156&1.768&0.923& 0.062&0.084 &0.2373 &1.4608\\
2.2 & 2.0412 &-0.3523&1.703&0.884& 0.056&0.073 &0.2562 &1.4011\\
2.3 & 2.0525 &-0.3908&1.643&0.848& 0.049&0.064 &0.2742 &1.3395\\
2.4 & 2.0671 &-0.4312&1.588&0.816& 0.041&0.055 &0.2913 &1.2759\\
2.5 & 2.0854 &-0.4740&1.536&0.787& 0.033&0.047 &0.3074 &1.2100\\
2.6 & 2.1081 &-0.5196&1.489&0.760& 0.025&0.040 &0.3226 &1.1418\\
2.7 & 2.1359 &-0.5687&1.445&0.735& 0.016&0.034 &0.3370 &1.0709\\
2.8 & 2.1698 &-0.6219&1.404&0.712& 0.007&0.028 &0.3505 &0.9971\\
2.9 & 2.2113 &-0.6803&1.365&0.691&-0.002&0.023 &0.3635 &0.9197\\
3.0 & 2.2621 &-0.7454&1.328&0.671&-0.016&0.019 &0.3764 &0.8380\\
3.1 & 2.3250 &-0.8190&1.294&0.652&-0.021&0.015 &0.3905 &0.7504\\
3.2 & 2.4039 &-0.9038&1.261&0.634&-0.030&0.012 &0.4095 &0.6528\\
3.3 & 2.5044 &-1.0040&1.230&0.618&-0.038&0.009 &0.4653 &0.5127\\
3.4 & 2.6359 &-1.1259&1.200&0.602&-0.046&0.006 &0.4436$^c$&0.4436$^c$\\
3.5 & 2.8140 &-1.2804&1.171&0.587&-0.054&0.004 &0.3946$^c$&0.3946$^c$\\
3.6 & 3.0678 &-1.4869&1.144&0.572&-0.061&0.002 &0.3411$^c$&0.3411$^c$\\
3.7 & 3.4570 &-1.7849&1.116&0.558&-0.066&0.001 &0.2822$^c$&0.2822$^c$\\
3.8 & 4.0852 &-2.2303&1.087&0.544&-0.066&0.000 &0.2136$^c$&0.2136$^c$\\
\hline
\hline
\end{tabular}
\end{center}
\end{table}


\begin{thebibliography}{99}
\bibitem{footnote1}
Here so called Fisher renormalization \cite{Fisher'68} holds. It states
that the critical exponents of an annealed system are determined by
those of the corresponding pure one.
\bibitem{footnote3}
We have used the change of the notations
$(m+8)Du/6 \rightarrow u$, $(mn+8)Dv/6 \rightarrow v$ and
$(m+8)D\beta_u/6 \rightarrow \beta_u$,
$(mn+8)D\beta_v/6 \rightarrow \beta_v$ with D being one loop
integral $D~=~1/(2\pi)^d \int {\rm d}^dq (q^2+1)^{-1}$ in order to define
convenient numerical scale in which the first two coefficients of the
functions $\beta_u$ and $\beta_v$ are $-1$ and $+1$.
\end{thebibliography}
\end{document}